# Eu ions and copper selenide nanoparticles within silica sol-gel matrices

Valerij Gurin[1], Alexander Alexeenko[2],

[1]Physico-Chemical Research Institute, Belarusian State University, Minsk, Belarus;
[2]Gomel State Technical University, Gomel, Belarus

**Abstract**

The sol-gel technique was developed for preparation of silica glasses doped with europium and copper chalcogenide nanoparticles. A synthesis of copper compounds proceeded within the solid matrix at the different steps of the process, and features of final products can be controlled by the precursor composition and heat treatment conditions. $Eu^{3+}$-ions within the silica matrix serves as a luminescent probe sensitive to the environment and a tool for studies of energy transfer between different absorption and emission centers with efficient light absorption in nanoparticles and emission by $Eu^{3+}$.

**1. Introduction**

New glassy materials with complex composition is of great interest both for optical applications and understanding of physical phenomena in solids associated with features of nanometer-scale constituents. In particular, luminescent properties of materials with semiconductor nanoparticles are studied intensively last years in connection with the size-dependent semiconductor nanoparticle optics. They can have own active luminescent centers or another species can participate in the light energy exchange to provide an efficient emission. The glasses considered in the present work are an example of the latter type of materials. We study the photoluminescence of rare earth ions ($Eu^{3+}$) within the silica monoliths containing semiconductor nanoparticles with own unique optical features. Copper chalcogenide particles were shown recently [1] to possess the promising combination of interband transitions with energy determined by $E_g$ and the additional near-IR absorbtion band originated from intraband transitions in semiconductor.

Optically transparent matrixes based on silicon dioxide fabricated with the sol-gel technique provide the wide opportunity to vary composition and properties of the materials and regulate state of nanoparticles dispersed within matrices. A doping with transition metal ions and small semiconductor particles appears to be very successful to attain new optical and luminescent features of silica-based materials [2]. A formation of active centres consisting from ions, clusters and nanoparticles strongly dependent on preparation conditions of the silica matrix and doping method. This type of silica-based materials can be prepared both in the form of monolithic glasses of high optical quality and thin films deposited on a transparent substrate. We concern here the selected types of monolithic glasses those are of more practical interest for non-linear optical studies [3] and discover the remarkable variation in the luminescence behavior [4].

## 2. Experimental

The preparation procedure of the silica-based monolithic glasses has been used by us [2-5] for similar materials containing copper oxide, sulfide and selenide. It consists in modification of the conventional sol-gel process with the catalytic hydrolysis of silicon-organic compounds followed by polycondensation of R-Si-O- fragments accompanied by amorphous network structure from -Si-O-Si- bridges with the four-coordinated silicon.

The scheme below explains the main preparation steps. An initial precursor sol was prepared by mixing of tetraethoxysilane (TEOS), water, and HCl (mole ratio 1:25:0.05, respectively). At this step the doping (Cu, Eu or a double Cu-Eu resulting finally after the subsequent steps in the fabrication of the three types of samples under study) was carried out by addition of $Cu(NO_3)_2$ and/or $Eu(NO_3)_3$ in molar ratio 2.5:1 into sol. A finely dispersed aerosil A-300 ($SiO_2$ particles about 10 nm in diameter) was added to the sol in molar ratio $TEOS:SiO_2=1:1$ in order to avoid strong volume contraction under the subsequent gelation and drying. The mixing was done with ultrasonic dispergation, and a solid remnant was separated with centrifugation. The efficient gelation was attained by increasing pH up to 6-7 with ammonia solution. Gels were dried in air under humidity control, and the dried gels were heated, at the first up to 600°C for several hours, and the final heating proceeded with the temperature growth rate about 100°/h up to 1200°C. This heating regime removed completely organic remnants from TEOS, physically absorbed water and the principal part of water and hydroxyl groups obtained within the matrix in the result of polycondensation. The final products fabricated were transparent glasses of good optical quality and mechanical strength, coloured in the case of samples with copper compounds. The composition close to $Cu_2Se$ was determined for the copper selenide produced in the final glasses [5], however some variation in the stoichiometry is possible accompanied by essential changes in optical properties. For optical measurements samples were cut up to appropriate thickness and polished.

Absorption spectra were recorded with a Specord M40 device in the UV/visible range, emission and excitation spectra were obtained with a Fluoromax-2 spectrofluorimeter under room temperature (excitation by Xe-lamp).

## 3. Absorption and excitation spectra

The absorption spectra of the glasses under study doped and undoped with $Eu^{3+}$ (Fig. 1) discover no rich structure that could appear due to $Eu^{3+}$ since the concentration of europium is small, and only the principal transition can be surely observed with the peak at $\lambda = 394$ nm. The weak shoulder at $\lambda = 250$ nm presents usually in different doped and undoped silica monoliths, this band is not of interest within the present paper. The excitation spectra indicates clearly the occurrence of $Eu^{3+}$ ions, excitation lines of which are well developed only in the case of copper-free materials since a presence of copper selenide provides the strong absorption. The positions of these lines are consistent with those observed for europium in different matrices [6,7].

The glass with copper selenide nanoparticles reveals another form of the spectrum (Fig. 2) that comprises the fundamental absorption edge ($\lambda \sim 500$ nm) due to interband transitions in semiconductor and the intense band in the near IR range (a part of it shown only). The nature of this band was proposed due to intraband transitions between levels formed by the surface layer of the copper selenide particles [1].

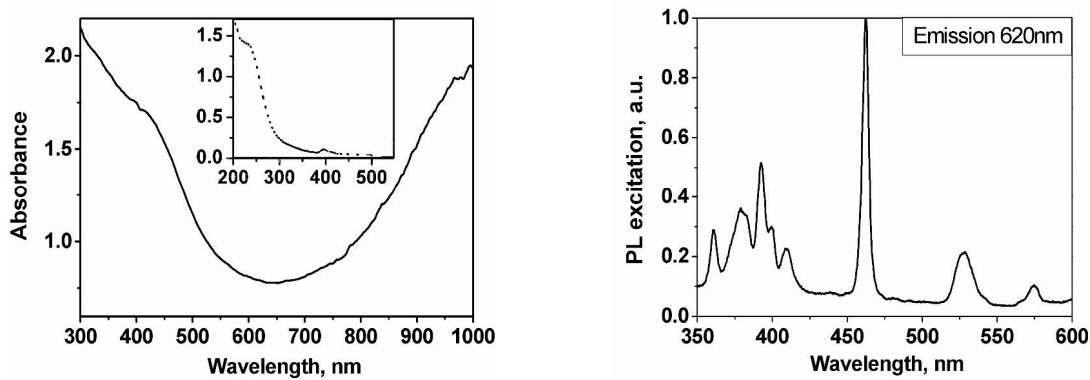

Fig. 1. Absorption spectra of the silica sol-gel glasses codoped with $Cu_2Se$ nanoparticles and $Eu^{3+}$ (main left plot) and these glasses without $Cu_2Se$ nanoparticles (inset) and excitation spectra of the 620 nm photoluminescence for the latter (right plot).

Transmission electron microscopy indicated the presence of nanoparticles with sizes 10-50 nm only for the copper-containing samples, Eu-doped materials without copper selenide shows no similar crystalline phase at the resolution used. The latter can be interpreted as highly dispersed state of europium (in the form of oxide) rather than nanoparticles those could be seen taking into account different transmittance of Eu and Si atoms for electron beam. Therefore, the doping mechanisms of amorphous silica with copper and europium are rather dissimilar and resulted to different phase state of the dopants. In the case of co-doped samples we can suppose that these nanoparticles are embedded into Eu-modified $SiO_2$ matrix. The contact of $Eu^{3+}$ ions with copper selenide nanoparticles easily can occur through the whole material providing efficient interaction via radiation absorbed and emitted.

## 4. Photoluminescence

The luminescence spectra (Fig. 2) have been collected for excitation with energy of photons higher than the principal absorption of $Eu^{3+}$ ($\lambda_{ex} = 354$ nm) and lower than this $Eu^{3+}$ absorption band ($\lambda_{ex} = 463$ nm). These two types of excitaions have distinguished the luminescence signals both from the $Eu^{3+}$-doped and $Eu^{3+}$-$Cu_2Se$-codoped samples. The $Eu^{3+}$-doped glass without copper showed no characteristic emission band under the short-wavelength excitation (Fig. 2, left). Nevertheless, the presence of europium ions in this sample has evidenced above by the excitation spectra (Fig. 1) and by the long-wavelength excitation as emission with the well resolved structure corresponding to the series of transitions $^5D_o \rightarrow {}^7F_j$ [6]. The amorphous silica matrix provides no any strong crystal field effects, and this luminescence behavior can argue on the location of europium within this glass as separated ions rather than any oxide nanocrystals. The present selection of excitation and emission conditions allows us to observe clearly the effect of copper selenide nanoparticles upon the europium luminescence that can be interpreted as an energy transfer between these two species embedded into the silica matrix. Fig. 2 (right) displays the luminescence spectra of the glass containing Eu and $Cu_2Se$

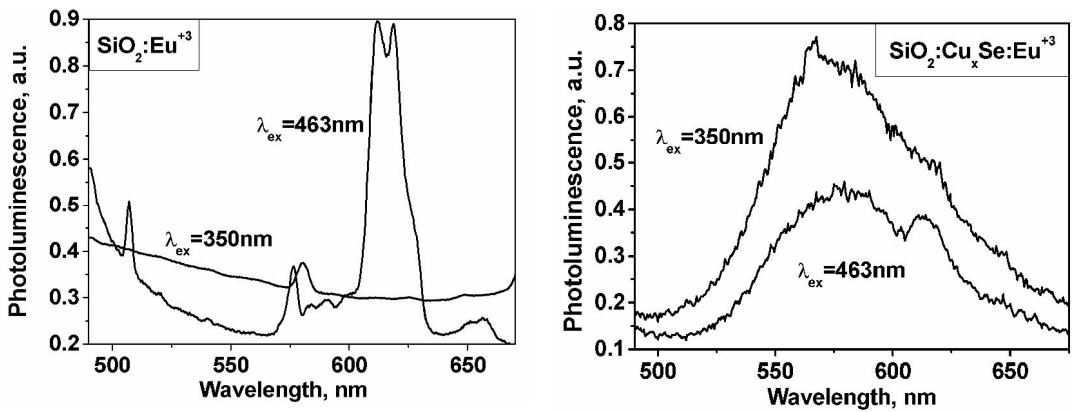

Fig. 2. Photolumionescence spectra of the silica sol-gel glasses doped with $Eu^{3+}$ (left) and codoped with $Cu_2Se$ nanoparticles and $Eu^{3+}$ (right) under excitation with different wavelength.

We observe similar spectra for excitation with short- and long-wavelength light. The first band (peaked at 570-575 nm) appears due to $Cu^+$ states those present in copper selenide particles. The additional band due to $Eu^{3+}$ is developed for the most intensive line of this ion on the background emission of copper, and this band is more intensive for $\lambda_{ex}$ = 463 nm. Thus, both luminescence centers ($Cu^+$ and $Eu^{3+}$) are active in this type of material with weak dependence on excitation wavelength. The presence of copper selenide nanoparticles influences emission europium resulting in its emission under different excitation wavelength.

## 5. Conclusions

New sol-gel silica glasses were fabricated with incorporation of copper selenide nanoparticles and europium, and their luminescence behavoir was studied. The effect of copper selenide nanoparticles upon the $Eu^{3+}$ luminescence is interpreted as the direct energy transfer between nanoparticles and europium localized within the glass matrix. The codoped material with $Cu_2Se$ nanoparticles in the contact with $Eu^{3+}$ ions provides the mixed emission of the both luminescence centers for different excitation wavelengths.

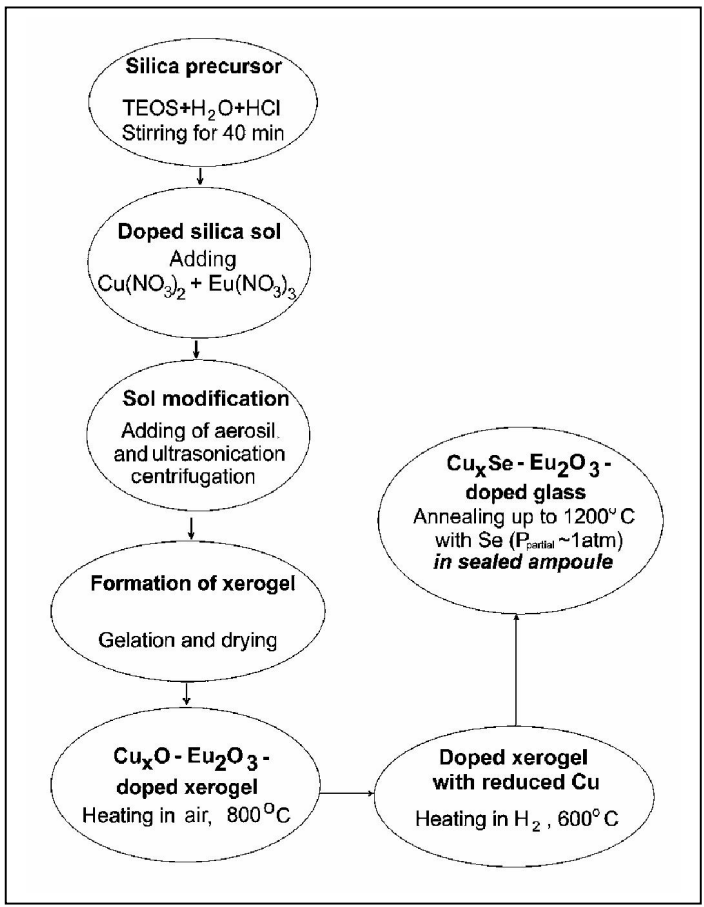